\documentclass{IEEEtran}

\usepackage{cite}
\usepackage{graphicx}
\usepackage{subfigure}
\usepackage{psfrag}
\usepackage{amsmath}
\usepackage{amssymb}

\newcommand{\mat}[1]{\ensuremath{\mathbf{#1}}}
\newcommand{\trans}{\ensuremath{^{\mathrm{T}}}}
\newcommand{\ctrans}{\ensuremath{^{\mathrm{H}}}}

\newcommand{\commpartm}{-\frac{\|\mat{y}_i-\sqrt{\mathcal{E}_d}h_i\mat{c}_m\|_2^2}{\sigma_c^2}}
\newcommand{\obserpart}{-\frac{(x-\theta)^2}{2\sigma_s^2}}
\newcommand{\commpart}{-\frac{\|\mat{y}_i-\sqrt{\mathcal{E}_d}h_i\mat{c}(x)\|_2^2}{\sigma_c^2}}

\newcommand{\figwidth}{3in}

\graphicspath{{figures/}}

\begin{document}

\title{Decentralized Estimation over Orthogonal Multiple-access Fading Channels
in Wireless Sensor Networks---Optimal and Suboptimal Estimators}
\author{Xin~Wang~and~Chenyang~Yang%
\thanks{This work was supported by the National Nature Science Foundation of China under Grant 60672103. %
Parts of this work were presented at IEEE Globecom'07, Washington, DC, United States, Nov. 2007.}%
\thanks{The authors are with Group 203, School of Electronics and Information Engineering, %
Beihang University, Beijing, 100191, China. Email: \emph{\{athody,cyyangbuaa\}@vip.sina.com}. %
Tel: +86-10-8231-7213 ext. 603/101. Fax: +86-10-8231-7213 ext. 201.}%
}
\maketitle

\begin{abstract}
Optimal and suboptimal decentralized estimators in wireless sensor
networks (WSNs) over orthogonal multiple-access fading channels are
studied in this paper. Considering multiple-bit quantization before
digital transmission, we develop maximum likelihood estimators
(MLEs) with both known and unknown channel state information (CSI).
When training symbols are available, we derive a MLE that is a
special case of the MLE with unknown CSI. It implicitly uses the
training symbols to estimate the channel coefficients and exploits
the estimated CSI in an optimal way. To reduce the computational
complexity, we propose suboptimal estimators. These estimators
exploit both signal and data level redundant information to improve
the estimation performance. The proposed MLEs reduce to traditional
fusion based or diversity based estimators when communications or
observations are perfect.
By introducing a general message
function, the proposed estimators can be applied when various
analog or digital transmission schemes are used.
The simulations show
that the estimators using digital communications with multiple-bit
quantization outperform the estimator using analog-and-forwarding
transmission in fading channels.
When considering the total bandwidth and energy
constraints, the MLE using
multiple-bit quantization is superior to that using binary
quantization at medium and high observation signal-to-noise ratio
levels.
\end{abstract}

\section{Introduction}\label{sec:orth:intro}

Wireless sensor networks (WSNs) consist of a number of sensors
 deployed in a field to collect information, e.g., measuring
physical parameters such as temperature and humidity. Since the
sensors are usually powered by batteries and have very limited
processing and communication abilities \cite{wsnsurvey}, the
parameters are often estimated in a decentralized way. In typical
WSNs for decentralized estimation, there exists a fusion center
(FC). The sensors transmit their locally processed observations to
the FC without inter-sensor communications, and the FC generates the
final estimation based on the received signals \cite{des06}.
\par

Both observation noise and communication error deteriorate the
performance of the decentralized estimation. Traditional
 fusion based estimators are able to minimize the mean
square error (MSE) of the parameter estimation by assuming perfect
communication links (see \cite{lxr03ef1} and references therein).
They reduce the observation noises by exploiting the redundant
observations provided by multiple sensors. However, their
performance will degrade dramatically when communication errors 
cannot be ignored or corrected. On the other hand, various wireless
communication technologies aimed at achieving transmission capacity
or improving reliability cannot necessarily minimize the MSE of the
parameter estimation. For example, although diversity combining
reduces the bit error rate (BER) of communications, it requires that
the signals transmitted from the sensors are identical, which is not
true in the context of WSNs due to the observation noises appeared
in sensors. This motivates the joint optimization of the
communication-oriented diversity combination and data
fusion-oriented estimator at the FC under realistic observation and
channel models, which uses MSE of the parameter estimation as the
performance metric.
\par

The bandwidth and energy constraints are two most important issues
that are addressed in WSNs. When strict bandwidth constraint is
taken into account, the decentralized estimation when the sensors
only transmit one bit (binary quantization) for each observation is
studied in
\cite{giannakis06bandwidth,giannakis06bandwidth2,zhiquanluo05,AysalBarner08Constrained,
lihb07dis,Fang2008}. Among them, \cite{giannakis06bandwidth},
\cite{giannakis06bandwidth2} introduce the maximum likelihood
estimation (MLE) and discuss the optimal quantization when the
communication channels are noiseless. Also considering noiseless
channels, \cite{zhiquanluo05} proposes a universal and isotropic
quantization rule, and \cite{lihb07dis,Fang2008} study the adaptive
binary quantization methods. When channels are noisy,
\cite{AysalBarner08Constrained} studies the MLE in additive white
Gaussian noise (AWGN) channels and introduces several low complexity
suboptimal estimators. It has been found that the binary
quantization is sufficient for decentralized estimation at low
observation signal-to-noise ratio (SNR), whereas the sensors need to
transmit a few extra bits when the observation SNR is high
\cite{giannakis06bandwidth}. When the energy constraint and general
multi-level quantizers are considered, the decentralized estimation
is studies under various channels. When communications are
error-free, quantization at the sensors are designed in
\cite{reibman93design, oppenheim01des, giannakis04energy,
jjxiao05decentralized,langtong06score}. The optimal trade-off
between the number of active sensors and the quantization bit-rate
of each sensor is investigated under total energy constraint in
\cite{Li2009distributed}. In binary symmetrical channels (BSCs), the
power scheduling is proposed to reduce the estimation MSE when the
best linear unbiased estimator (BLUE) and a quasi-BLUE,
where quantization noise is taken into account,
are used at the FC \cite{goldsmith05, jinjun06power}.
To the best of the authors' knowledge, the optimal decentralized
estimator using multiple-bit quantization in fading channels is
still not available. Although the MLE proposed in AWGN channel
\cite{AysalBarner08Constrained} can be applied for fading channels
if the channel state information (CSI) is known at the FC, it only
considers binary quantization.
\par

Besides the decentralized estimation based on digital
communications, the estimation based on analog communications
receives considerable attentions due to the important conclusions
drawn from the studies in multi-terminal coding problem
\cite{gastparphd, sourcechannel03vetterli}. The most popular scheme
is amplify-and-forward (AF) transmission, which is proved to be
optimal in quadratic Gaussian sensor networks
with AWGN multiple-access channels (MACs) \cite{gastpar07uncoded}.
The power scheduling and energy efficiency of the AF transmission
are studied under AWGN channels in \cite{goldsmith05energy} and
\cite{goldsmith05powerefficient}. It shows that the AF transmission
is more energy-efficient than that of digital communications with
certain coding and modulation schemes. In fading channels, the AF
transmission is not optimal any more both in orthogonal MACs
\cite{zhiquanluo07multiterminal,esdiversity07cui, Bai2009outage} and
in non-orthogonal MACs \cite{sayeed05optimal}. The outage laws of
the estimation diversity with AF transmission in fading channels are
studied in \cite{esdiversity07cui} and \cite{Bai2009outage} in
different asymptotic regimes. These studies, especially the results
in \cite{zhiquanluo07multiterminal}, indicate that the separate
source-channel coding scheme outperforms the AF transmission, which
is a simple joint source-channel coding scheme, and is actually
optimal in fading channels with orthogonal multiple-access
protocols.
\par

In this paper, we develop the optimal and suboptimal decentralized
estimators for a deterministic parameter considering digital
communication systems. The observations of sensors are quantized,
coded and modulated, then transmitted to the FC with orthogonal MACs
over Rayleigh fading channels. Uniform quantization is used since it
is optimal for deterministic parameters. Because the binary
quantization is only applicable for low observation SNR
levels\cite{giannakis06bandwidth,Li2009distributed}, a general multi-bit quantizer is
considered.
\par

We strive for deriving the MLE and feasible suboptimal estimator
when different local processing and communication strategies are
used. To this end, we first present a general message function to
represent various quantization and transmission schemes. We then
derive the MLE for an unknown parameter with known CSI at the FC. In
typical WSNs, the sensors usually cannot transmit too many training
symbols for the receiver to estimate or track channel coefficients
due to both energy and bandwidth constraints. Therefore, we will
also consider the case that the CSI is unknown at the FC when no or
only a few training symbols are available, which is of practical
significance. In order to reduce the computational complexity, we
will introduce suboptimal estimators following the hint provided by
the structure of the MLEs.
\par

Our contributions are summarized as follows. We develop the
decentralized MLEs with known and unknown CSI at the FC over
orthogonal MACs with Rayleigh fading.
The performance of the MLEs serves as the practical performance
lower bounds of the decentralized estimation in orthogonal MACs. To
provide feasible estimators with affordable complexity, we propose a
suboptimal algorithm, which can be viewed as modified
expectation-maximization (EM) algorithm \cite{emalgo77}. By studying
the special cases for  error-free communications or noiseless
observations, we show that the MLEs degenerate into the well-known
centralized fusion estimator---BLUE, or the maximal ratio combiner
(MRC) based estimator when CSI is known and a subspace based
estimator when CSI is unknown. This indicates that our estimators
can exploit both data level redundancy and signal level redundancy
provided by the multiple sensors when both observation noises and
communication errors are present. By introducing a general
message function that can describe various quantization and
transmission schemes, the proposed decentralized estimators can also
be applied for the WSNs where AF transmission or digital
transmission with binary quantization are used. Therefore, our
estimators can bridge the gap between the estimators using two
extreme case quantization.
\par

The rest of the paper is organized as follows. Section
\ref{sec:orth:model} describes the system models we considered.
Section \ref{sec:orth:opt} presents the MLEs with known and unknown
CSI, and Section \ref{sec:orth:subopt} introduces the suboptimal
estimators. In Section \ref{sec:specialcases}, we analyze several
special cases of the MLEs. In Section \ref{sec:DISCUSS}, we discuss
the codebook design issue, the computational complexity, and the
asymptotic performance of the presented MLEs. Simulations are
provided in Section \ref{sec:orth:simulation}, and the conclusions
are given in Section \ref{sec:orth:con}.

\section{System Model}\label{sec:orth:model}

We consider a typical kind of WSNs that consists of $N$ sensors and
a FC to measure an unknown deterministic parameter $\theta$, where
there are no inter-sensor communications among the sensors. The
sensors transmit their quantized observations to the FC over
Rayleigh fading channels. Assume that the sensors use ideal
orthogonal multiple-access protocols, such as TDMA and FDMA, to transmit their signals to
the FC. Then the FC can separate the received signals from different
sensors without inducing interference.
\par

\begin{figure}[thp]
    \begin{center}
        \includegraphics[width=\figwidth]{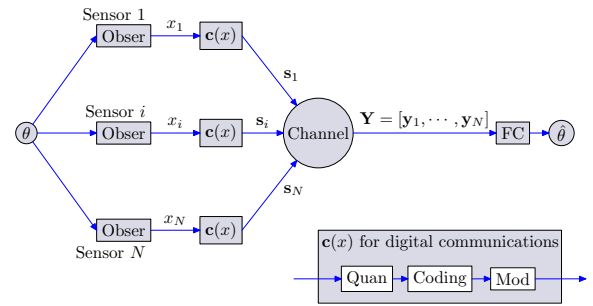}
    \end{center}
    \caption{The diagram of the decentralized estimation system we considered.}
    \label{fig:orth:model}
\end{figure}

Figure \ref{fig:orth:model} is a diagram of the decentralized
estimation system considered. The sensors process their observations
for the parameter $\theta$ before transmission. For digital
communications, the processing includes quantization, channel coding
and modulation, etc. For analog communications, the processing may be
simply amplification before transmission. A function $\mat{c}(x)$,
named as messaging function, is used to describe the local processing
for both digital and analog communication systems. The transmission
signals of the sensors arrive at the FC through independent Rayleigh
fading channels, and the FC uses the received signals to estimate
$\theta$.

\par

In the subsequent sections, we will first derive the decentralized
estimators considering digital communication, then extend the
results to the case using analog communication.

\subsection{Observation Model}\label{sec:orth:model:obser}

The observation for the unknown parameter $\theta$ provided by the $i$-th sensor is,
\begin{equation}
x_i=\theta+n_{s,i},\,\,i=1,\cdots,N, \label{eq:obser}
\end{equation}
where $n_{s,i}$ is the independent identically distributed (i.i.d.)
Gaussian observation noise with zero mean and variance $\sigma_s^2$,
and $\theta$ is bounded within a dynamic range $[-V, +V]$.

\subsection{Quantization, Coding and Modulation}\label{sec:orth:model:localproc}

We use the messaging function
$\mat{c}(x)|\mathbb{R}\rightarrow\mathbb{C}^L$, which maps the
observations to transmission symbols, to represent all the
processing at the sensors including quantization, coding and
modulation. To facilitate the analysis, the energy of the transmission symbols
is normalized to 1,
\begin{equation}
   \mat{c}(x)\ctrans\mat{c}(x)=1,\quad\forall x\in\mathbb{R}.
   \label{eq:nmenergynorm}
\end{equation}
\par

We consider the uniform quantization, which is optimal for deterministic
parameters or for parameters with unknown statistics. For an
$M$-level uniform quantizer, define the granular region of the
quantizer as $[-W,+W]$, then all the possible quantized values of the
observations can be written as,
\begin{equation}
    S_m=m\Delta - W, \quad m=0,\cdots,M-1,
   \label{eq:nmsm}
\end{equation}
where $\Delta = 2W/(M-1)$ is the quantization interval.
\par

The observations are rounded to the nearest $S_m$, therefore
$\mat{c}(x)$ is a piecewise constant function described
as,
\begin{equation}
    \mat{c}(x)=\left\{
    \begin{array}{l c}
        \mat{c}_0, & -\infty < x \leq S_0+\frac{\Delta}{2}\\
        \mat{c}_m, & S_m-\frac{\Delta}{2} < x \leq S_m+\frac{\Delta}{2}\\
        \mat{c}_{M-1}, & S_{M-1}-\frac{\Delta}{2} < x < +\infty
    \end{array}
    \right. ,
    \label{eq:nmcm}
\end{equation}
where $\mat{c}_m=[c_{m,1},\cdots,c_{m,L}]\trans$ is the $L$ symbols
corresponding to the quantized observation $S_m$ to be transmitted,
$m=0,\cdots,M-1$.
\par

Under the assumption that $W$ is much larger than the dynamic range of $\theta$,
the probability that
$|x_i|>W$ can be ignored. 
Then $\mat{c}(x)$ is simplified as,
\begin{equation}
   \mat{c}(x)=\mat{c}_m, \quad S_m-\frac{\Delta}{2} < x \leq S_m+\frac{\Delta}{2}.
   \label{eq:nmcmapprox}
\end{equation}
\par

Define the transmission codebook as,
\begin{equation}
    \mat{C}_t=[\mat{c}_0,\cdots,\mat{c}_{M-1}]\in\mathbb{C}^{L\times M},
   \label{eq:nmcodebook}
\end{equation}
%
%
which can be used to describe any coding and modulation scheme
following the $M$-level quantization.
\par

The sensors can use various codes such as natural binary codes to
represent the quantized observations. Since the focus of this paper
is to design the decentralized estimators, we will not optimize the
transmission codebook for the parameter estimation.

\subsection{Received Signals}\label{sec:models:rxsignal}

Since we consider orthogonal MACs, we assume that the FC can
perfectly separate and synchronize to the received signals from
different sensors. Assume that the channels are block fading, i.e.,
the channel coefficients are invariant during the period that
sensors transmit $L$ symbols representing one observation.
After matched filtering and symbol-rate sampling, the $L$ received
samples corresponding to the $L$ transmitted symbols from the $i$-th
sensor can be expressed as,
\begin{equation}
    \mat{y}_i=\sqrt{\mathcal{E}_d}h_i\mat{c}(x_i)+\mat{n}_{c,i},\, i=1,\cdots,N,
    \label{eq:nmchannel}
\end{equation}
where $\mat{y}_i=[y_{i,1},\cdots,y_{i,L}]^{\mathrm{T}}$, $h_i$ is
the channel coefficient subjecting to complex Gaussian distribution
with zero mean and unit variance,
$\mat{n}_{c,i}$ is a vector of thermal noise at the receiver
subjecting to complex Gaussian distribution with zero mean and
covariance matrix $\sigma_c^2\mat{I}$, and $\mathcal{E}_d$ is the
transmission energy for each observation.
\par

\section{Optimal Estimators with or without CSI}\label{sec:orth:opt}

In this section, we derive the MLEs when CSI is known or
unknown at the receiver of the FC, respectively. The MLE
using training symbols in the transmission codebook is also studied as a
special form of the MLE with unknown CSI.

\subsection{MLE with Known CSI}\label{sec:orth:mle}

Given $\theta$, the received signals from different sensors are
statistically independent.
If the CSI is known at the receiver of the FC, the log-likelihood
function is,
{\setlength{\arraycolsep}{1pt}
\begin{eqnarray}
    \log p(\mat{Y}|\mat{h},\theta)&=& \sum_{i=1}^N\log p(\mat{y}_i|h_i,\theta)\nonumber\\
    &=& \sum_{i=1}^N \log\left( \int_{-\infty}^{+\infty}p(\mat{y}_i|h_i,x)p(x|\theta)\mathrm{d}x \right)
    \label{eq:lfsecond}
\end{eqnarray}}
where $\mat{h}=[h_1,\cdots,h_N]\trans$ is the channel coefficients vector, and
%
$p(x|\theta)$ is the conditional probability density function (PDF)
of the observation given $\theta$. Following the observation model
shown in (\ref{eq:obser}), we have,
\begin{equation}
    p(x|\theta)=\frac{1}{\sqrt{2\pi}\sigma_s}\exp\left( -\frac{(x-\theta)^2}{2\sigma_s^2} \right).
    \label{eq:pxitheta}
\end{equation}
\par

According to the received signal model shown in (\ref{eq:nmchannel}),
the PDF of the received signals given CSI and the observation of the
sensors is,
\begin{equation}
    p(\mat{y}_i|h_i,x)=\frac{1}{(\pi\sigma_c^2)^L}\exp\left(
    -\frac{\|\mat{y}_i-\sqrt{\mathcal{E}_d}h_i\mat{c}(x)\|_2^2}{\sigma_c^2} \right),
    \label{eq:pyixi}
\end{equation}
where $\|\mat{z}\|_2=(\mat{z}\ctrans\mat{z})^{1/2}$ is $l_2$ norm of vector $\mat{z}$.
\par

Substitute (\ref{eq:pxitheta}) and (\ref{eq:pyixi}) to (\ref{eq:lfsecond}),
then the log-likelihood function becomes,
{\setlength{\arraycolsep}{1pt}
\begin{eqnarray}
    \log p(\mat{Y}|\mat{h},\theta)&=& \sum_{i=1}^{N}\log\left(
    \int_{-\infty}^{+\infty}\exp\left( -\frac{(x-\theta)^2}{2\sigma_s^2}-
    \right.\right.\nonumber \\ && \left.\left.
    \frac{\|y_i-\sqrt{\mathcal{E}_d}h_i\mat{c}(x)\|_2^2}{\sigma_c^2}\right)
    \right)+a,
    \label{eq:lffinal}
\end{eqnarray}}
where $a =
\log\left(\frac{1}{\sqrt{2}\pi^{L+\frac{1}{2}}\sigma_s\sigma_c^{2L}}\right)$
is a constant that does not affect the estimation.
From now on, we will omit the constant 
when we write likelihood functions for simplicity.
\par

Now we consider the form of the likelihood function for digital
communications, where $\mat{c}(x)$ is a piecewise constant function
described in (\ref{eq:nmcmapprox}). Substituting (\ref{eq:nmcmapprox})
to (\ref{eq:lffinal}), we have,
\begin{equation}
    \log p(\mat{Y}|\mat{h},\theta)=\sum_{i=1}^{N}\log\left(\sum_{m=0}^{M-1}
    p(\mat{y}_i|h_i,\mat{c}_m)p(S_m|\theta)
    \right),
    \label{eq:lfd}
\end{equation}
where $p(\mat{y}_i|h_i,\mat{c}_m)$ is the PDF of the received
signals given CSI and the transmitted symbols of the sensors, which
is,
\begin{eqnarray}
    {p}(\mat{y}_i|h_i,\mat{c}_m)=\frac{1}{(\pi\sigma_c^2)^L}\exp\left(
    \commpartm
    \right),
    \label{eq:defpyihicm}
\end{eqnarray}
and $p(S_m|\theta)$ is the probability mass function (PMF) of the quantized observation given $\theta$,
which is,
\begin{eqnarray}
    {p}(S_m|\theta)=\mathrm{Q}\left( \frac{S_m-\frac{\Delta}{2}-\theta}{\sigma_s}\right)-
    \mathrm{Q}\left(\frac{S_m+\frac{\Delta}{2}-\theta}{\sigma_s} \right),
    \label{eq:defpsm}
\end{eqnarray}
where
$\mathrm{Q}(x)=\frac{1}{\sqrt{2\pi}}\int_{x}^{\infty}\exp\left( -\frac{t^2}{2}\right)\mathrm{d}t$.
\par

The MLE that maximizes (\ref{eq:lfd}) is then,
{\setlength{\arraycolsep}{1pt}
\begin{eqnarray}
    \hat{\theta}&=& \arg\max_{\theta} \sum_{i=1}^N\log\left(
    \sum_{m=0}^{M-1}
    \exp\left(
    -\frac{\|\mat{y}_i-\sqrt{\mathcal{E}_d}h_i\mat{c}_m\|_2^2}{\sigma_c^2}
    \right)\right. \nonumber \\ && \left.
    \left(
    \mathrm{Q}\left( \frac{S_m-\frac{\Delta}{2}-\theta}{\sigma_s} \right) -
    \mathrm{Q}\left( \frac{S_m+\frac{\Delta}{2}-\theta}{\sigma_s} \right)
    \right)
    \right).
    \label{eq:mle}
\end{eqnarray}}
\par



The log-likelihood function in (\ref{eq:mle}) is non-concave and has
multiple extrema. It is difficult to find a closed-form expression
of $\hat{\theta}$ or to compute $\hat{\theta}$ using high efficient
numerical methods.
\par


\subsection{MLE with Unknown CSI}\label{sec:orth:mlenocsi}

When the CSI is unknown at the FC, the log-likelihood function is,
\begin{equation}
  \log p(\mat{Y}|\theta) = \sum_{i=1}^N\left(
  \log\left(
  \int_{-\infty}^{+\infty}p(\mat{y}_i|x)p(x|\theta)\mathrm{d}x
  \right)
  \right),
  \label{eq:lfnocsifirst}
\end{equation}
which has a similar form to the likelihood function with known CSI
shown in (\ref{eq:lfsecond}).
\par

According to the received signal model shown in (\ref{eq:nmchannel}),
given $x$, $\mat{y}_i$ subjects to zero mean complex Gaussian
distribution, i.e.,
\begin{equation}
    p(\mat{y}_i|x)=\frac{1}{\pi^{L}\det\mat{R}_{\mat{y}}}\exp\left(
    -\mat{y}_i\ctrans\mat{R}_{\mat{y}}^{-1}\mat{y}_i
    \right),
    \label{eq:pdfyi}
\end{equation}
where $\mat{R}_{\mat{y}}$ is the covariance matrix of $\mat{y}_i$,
which is,
\begin{equation}
    \mat{R}_{\mat{y}}=\sigma_c^2\mat{I}+\mathcal{E}_d\mat{c}(x)\mat{c}(x)\ctrans.
    \label{eq:corrvaryi}
\end{equation}
\par

It is readily to find that one eigenvalue of $\mat{R}_{\mat{y}}$
equals to $\mathcal{E}_d+\sigma_c^2$, and all other eigenvalues
equal to $\sigma_c^2$. Thereby the determinant of
$\mat{R}_{\mat{y}}$ is,
\begin{equation}
    \det\mat{R}_{\mat{y}}=(\mathcal{E}_d+\sigma_c^2)\sigma_c^{2(L-1)}.
    \label{eq:detryi}
\end{equation}
\par

Following the Matrix Inverse Lemma, we have,
\begin{equation}
    \mat{R}_{\mat{y}}^{-1}=\frac{1}{\sigma_c^2}\mat{I}-
    \frac{\mathcal{E}_d}{\sigma_c^2(\mathcal{E}_d+\sigma_c^2)}\mat{c}(x)\mat{c}(x)\ctrans.
    \label{eq:invryi}
\end{equation}
\par

Substituting (\ref{eq:detryi}) and (\ref{eq:invryi}) to (\ref{eq:pdfyi}), $p(\mat{y}_i|x)$ becomes,
{\setlength{\arraycolsep}{1pt}
\begin{eqnarray}
    p(\mat{y}_i|x)&=& \alpha\exp\left(
    -\frac{\|\mat{y}_i\|_2^2}{\sigma_c^2}+
    \frac{\mathcal{E}_d\mat{y}_i\ctrans\mat{c}(x)\mat{c}(x)\ctrans\mat{y}_i}
    {\sigma_c^2(\mathcal{E}_d+\sigma_c^2)}
    \right)\nonumber\\
    &=& \alpha\exp\left(
    -\frac{\|\mat{y}_i\|_2^2}{\sigma_c^2}+
    \frac{\mathcal{E}_d|\mat{y}_i\ctrans\mat{c}(x)|^2}{\sigma_c^2(\mathcal{E}_d+\sigma_c^2)}
    \right),
    \label{eq:pdfyifinal}
\end{eqnarray}}
where $\alpha=\frac{1}{\pi^{L}(\mathcal{E}_d+\sigma_c^2)\sigma_c^{2(L-1)}}$ is a constant.
\par

%

Upon substituting (\ref{eq:pdfyifinal}) and (\ref{eq:defpsm}) to
(\ref{eq:lfnocsifirst}), the log-likelihood function becomes,
{\setlength{\arraycolsep}{1pt}
\begin{eqnarray}
    \log p(\mat{Y}|\theta)&=&
    \sum_{i=1}^N\log\left(
    \int_{-\infty}^{+\infty}\exp\left( \obserpart-
    \right.\right.\nonumber \\ && \left.\left.
    \frac{\|\mat{y}_i\|_2^2}{\sigma_c^2}+
    \frac{\mathcal{E}_d|\mat{y}_i\ctrans\mat{c}(x)|^2}{\sigma_c^2(\mathcal{E}_d+\sigma_c^2)}
    \right)\mathrm{d}x
    \right).
    \label{eq:lfnocsi}
\end{eqnarray}}
Then, the MLE is obtained as,
\begin{equation}
  \hat{\theta}=\arg\max_{\theta}\log p(\mat{Y}|\theta).
   \label{eq:mlenocsi}
\end{equation}
\par

\subsection{MLE with Unknown CSI using Training Symbols}

In typical communication systems, the transmitted symbols may
consist of training symbols to facilitate channel estimation. In
this subsection, we will analyze the MLE for such
transmission schemes.
\par

Define $\mat{c}_p$ as a vector which consists of $L_p$ training
symbols. Each transmission for an observation will begin with
transmitting $\mat{c}_p$, then transmitting the data symbols defined as $\mat{c}_d(x)$.
Thus the messaging function becomes,
\begin{equation}
    \mat{c}(x)=\left(
    \begin{array}{c}
        \mat{c}_p\\
        \mat{c}_d(x)
    \end{array}
    \right).
    \label{eq:defcxpilot}
\end{equation}
\par

Upon substituting
this expression to (\ref{eq:lfnocsi}) and ignoring all the terms
that does not affect the estimation, we obtain the likelihood
function as,
{\setlength{\arraycolsep}{1pt}
\begin{eqnarray}
    \log p(\mat{Y}|\theta)&=&  
    \sum_{i=1}^N\log\left(
    \int_{-\infty}^{+\infty}\exp\left( \obserpart +
    \right.\right.\nonumber\\ &&
    \beta|\mat{y}_{i,d}\ctrans\mat{c}_d(x)|^2 +
    2\beta\Re\{\mat{c}_p\ctrans\mat{y}_{i,p}\mat{y}_{i,d}\ctrans\mat{c}_d(x)\}
    \Big)\mathrm{d}x
    \bigg),\nonumber\\
    \label{eq:lfnocsisemiblind}
\end{eqnarray}}
where 
$\mat{y}_{i,p}$ and $\mat{y}_{i,d}$ are the received
signals corresponding to the training symbols and data symbols,
respectively, and 
$\beta=\mathcal{E}_d/\sigma_c^2(\mathcal{E}_d+\sigma_c^2)$ is a constant.

\par

Now we show that $\mat{c}_p\ctrans\mat{y}_{i,p}$ in
(\ref{eq:lfnocsisemiblind}) can be regarded as the minimum mean
square error (MMSE) estimate for the channel coefficient $h_i$
except for a constant factor. Since both $h_i$ and the receiver
thermal noise are complex Gaussian distributed, the MMSE estimate of
$h_i$ is equivalent to its linear MMSE estimate, which is,
\begin{equation}
    \hat{h}_i=(\mat{R}_{y_p}^{-1}\mat{r}_{yh})\ctrans\mat{y}_{i,p},
    \label{eq:esthi}
\end{equation}
where $\mat{r}_{yh}=\mathrm{E}[\mat{y}_{i,p}h_i^*]$, and
$\mat{R}_{y_p}$ is the covariance matrix of $\mat{y}_{i,p}$.

According to the received signal model, we have
\begin{eqnarray}
    \mat{r}_{yh}=\mathrm{E}[(\sqrt{\mathcal{E}_d}h_i\mat{c}_p+\mat{n}_{ci,p})h_i^*]=
    \sqrt{\mathcal{E}_d}\mat{c}_p,
    \label{eq:esthiryh}
\end{eqnarray}
and
\begin{eqnarray}
    \mat{R}_{y_p}=\mathcal{E}_d\mat{c}_p\mat{c}_p\ctrans+\sigma_c^2\mat{I}.
    \label{eq:esthiryp}
\end{eqnarray}
\par

With the Matrix Inverse Lemma, $\mat{R}_{y_p}^{-1}$ is expressed as, 
\begin{eqnarray}
    \mat{R}_{y_p}^{-1}&=& \frac{1}{\sigma_c^2}\left(
    \mat{I}-\frac{\mathcal{E}_d\mat{c}_p\mat{c}_p\ctrans}
    {\sigma_c^2+\mathcal{E}_d\mat{c}_p\ctrans\mat{c}_p}
    \right).
    \label{eq:esthrypinv}
\end{eqnarray}
\par

Upon substituting (\ref{eq:esthiryp}) and (\ref{eq:esthrypinv}) to
(\ref{eq:esthi}), the MMSE channel estimate becomes,
\begin{equation}
    \hat{h}_i=\frac{\sqrt{\mathcal{E}_d}L}{L\sigma_c^2+L_p\mathcal{E}_d}\mat{c}_p\ctrans\mat{y}_{i,p}.
    \label{eq:esthfinal}
\end{equation}
\par

Let $\kappa =\frac{\sqrt{\mathcal{E}_d}L}{L\sigma_c^2+L_p\mathcal{E}_d}$, then we have
$\mat{c}_p\ctrans\mat{y}_{i,p}=\hat{h}_i/\kappa$.
Substituting it to (\ref{eq:lfnocsisemiblind}), we obtain,
{\setlength{\arraycolsep}{1pt}
\begin{eqnarray}
    \log p(\mat{Y}|\theta)&=& 
    \sum_{i=1}^N\log\left(
    \int_{-\infty}^{+\infty}\exp\left( \obserpart +
    \right.\right.\nonumber \\ && \left.\left.
    \beta|\mat{y}_{i,d}\ctrans\mat{c}_d(x)|^2 +
    \frac{2\beta}{\kappa}\Re\{\hat{h}_i\mat{y}_{i,d}\ctrans\mat{c}_d(x)\}
    \right)\mathrm{d}x
    \right).\nonumber\\
    \label{eq:lfsemifinal}
\end{eqnarray}}
\par

In the following, we will show that the MLE in this case is
equivalent to a two-stage estimator. During the first stage, the FC
uses (\ref{eq:esthfinal}) to obtain the MMSE estimate of $h_i$. It
can be modeled as $\hat{h}_i={h}_i+\epsilon_{h_i}$, where
$\epsilon_{h_i}$ is the estimation error subjecting to complex
Gaussian distributed with zero mean and variance
$L\sigma_c^2/(L\sigma_c^2+L_p\mathcal{E}_d)$. During the second
stage, the FC conducts the MLE using $\hat{h}_i$.
\par

Substituting $\hat{h}_i$ to (\ref{eq:nmchannel}), the received
signal of the data symbols becomes,
\begin{equation}
    \mat{y}_{i,d}=\sqrt{\mathcal{E}_d}\hat{h}_i\mat{c}_d(x)-\sqrt{\mathcal{E}_d}\epsilon_{h_i}
    \mat{c}_d(x) + \mat{n}_{ci,d},
    \label{eq:modelyid}
\end{equation}
where $\mat{n}_{ci,d}$ is the receiver thermal noise.
\par

Deriving the conditional PDF $p(\mat{y}_{i,d}|\hat{h}_i, x)$ with
(\ref{eq:modelyid}), we can obtain a likelihood function which is
exactly the same as (\ref{eq:lfsemifinal}).
This implies that the MLE with unknown CSI can exploit the available
training symbols implicitly to provide the optimal channel
estimation and then use it to provide the optimal estimation of
$\theta$.
\par

Note that the likelihood function in (\ref{eq:lfsemifinal}) is
different from the likelihood function that uses the estimated CSI
as the true value of the channel coefficients, which is,
{\setlength{\arraycolsep}{1pt}
\begin{eqnarray}
    \log&& p(\mat{Y}|h_i=\hat{h}_i, \theta)=
    \sum_{i=1}^N\log\left(
    \int_{-\infty}^{+\infty}\exp\left( \obserpart +
    \right.\right.\nonumber\\ && \left.\left.
    \frac{2\sqrt{\mathcal{E}_d}}{\sigma_c^2}\Re\{\hat{h}_i\mat{y}_{i,d}\ctrans\mat{c}_d(x)\}
    \right)\mathrm{d}x
    \right).
    \label{eq:lfeshfinal}
\end{eqnarray}}
This is a coherent estimator. By contrast, there exist both a
coherent term $\Re\{\hat{h}_i\mat{y}_{i,d}\ctrans\mat{c}_d(x)\}$ and
a non-coherent term $|\mat{y}_{i,d}\ctrans\mat{c}_d(x)|^2$ in
(\ref{eq:lfsemifinal}). This means that the MLE shown in
(\ref{eq:lfsemifinal}) uses the channel estimate as ``partial'' CSI
after accounting for the channel estimation errors. The true value
of the channel coefficients contained in the channel estimation
corresponds to the coherent term of the log-likelihood function,
whereas the uncertainty in the channel estimation, i.e., the
estimation errors, leads to the non-coherent term. We will compare
the performance of the two estimators through simulations in Section
\ref{sec:orth:simulation}.
\par


\section{Suboptimal Estimator}\label{sec:orth:subopt}

In the previous section, we developed the optimal estimators for the
considered decentralized estimation systems, which are not feasible
for practical systems due to their prohibitive computational
complexity. Nevertheless, their performance can serve as the
practical upper-bound when both observation noise and communication
error are present, and their structure provides us some hint to
derive low complexity suboptimal estimators. In this section, we
take the suboptimal estimator with known CSI as an example. The
estimator with unknown CSI can be obtained following the same
principle.
\par

We first consider an approximation of the PMF $p(S_m|\theta)$.
Following the Lagrange Mean Value Theorem,  there exists $\xi$ in an
interval $[\frac{S_m-\frac{\Delta}{2}-\theta}{\sigma_s},
\frac{S_m+\frac{\Delta}{2}-\theta}{\sigma_s}]$ that satisfies,
\begin{eqnarray}
    p(S_m|\theta)=
    -Q'(\xi)\frac{\Delta}{\sigma_s}
    =\frac{\Delta}{\sqrt{2\pi}\sigma_s}\exp\left( -\frac{\xi^2}{2} \right).
    \label{eq:midvalue}
\end{eqnarray}
\par

If the quantization interval $\Delta$ is small enough, we can let
$\xi$ equal to the middle value of the interval, i.e.,
$\xi=(S_m-\theta)/\sigma_s$, and obtain an approximate expression of
the PMF as,
\begin{equation}
    p(S_m|\theta)\approx p_A(S_m|\theta)\triangleq
    \frac{\Delta}{\sqrt{2\pi}\sigma_s}\exp\left( -\frac{(S_m-\theta)^2}{2\sigma_s^2} \right).
    \label{eq:psmtheta}
\end{equation}
\par


Substituting (\ref{eq:psmtheta}) to (\ref{eq:lfd}) and computing its
partial derivative with respect to $\theta$, the likelihood
equation is simplified as,
\begin{eqnarray}
    \theta
    = \frac{1}{N}\sum_{i=1}^N\left(
    \frac{\displaystyle\sum_{m=0}^{M-1}p(\mat{y}_i|h_i,\mat{c}_m)p_A(S_m|\theta)S_m}
    {\displaystyle\sum_{m=0}^{M-1}p(\mat{y}_i|h_i,\mat{c}_m)p_A(S_m|\theta)}
    \right),
    \label{eq:lfeqn}
\end{eqnarray}
which is the necessary condition for the MLE of $\theta$.
\par

Unfortunately, we cannot obtain an explicit estimator for $\theta$
from this equation because the right hand side of the likelihood
equation also contains $\theta$.
However, considering the property of the conditional PDF, (\ref{eq:lfeqn}) becomes,
\begin{eqnarray}
    \theta
    &=& \frac{1}{N}\sum_{i=1}^N\left( \sum_{m=0}^{M-1}p(S_{m}|\mat{y}_i,h_i,\theta)S_{m} \right)
    \nonumber\\
    &=& \frac{1}{N}\sum_{i=1}^{N} \mathrm{E}\left[ S_{m}|\mat{y}_i,h_i,\theta \right].
    \label{eq:lfeqn2}
\end{eqnarray}
\par

If we assume that $\theta$ in the right hand side of the likelihood
equation is known, the right hand side of (\ref{eq:lfeqn2}) is
actually the MMSE estimator of $S_{m_i}$, i.e.,
$\hat{S}_{m_i}=\mathrm{E}\left[ S_{m_i}|y_i,h_i,\theta \right]$.
This indicates that the MLE can be regarded as a two stage
estimator. During the first stage, it estimates $S_{m_i}$,
$i=1,\cdots,N$, with the received signals from each sensor. During
the second stage, it combines $\hat{S}_{m_i}$ by a sample mean
estimator. These two stages are consistent with the two steps of the
EM algorithm \cite{emalgo77}.
The first stage is the expectation step (E-step) and the second
stage is the maximization step (M-step) of the algorithm. The
set of the quantized observations $S_{m_i}$, which is the sufficient
statistics for estimating $\theta$, is the complete data of the EM
algorithm.
\par

We present a suboptimal estimator with a similar two-stage
structure. This estimator can be viewed as a modified EM algorithm.
Because the likelihood function shown in (\ref{eq:lfd}) has multiple
extrema and the equation shown in (\ref{eq:lfeqn}) is only a
necessary condition, the initial value of the iterative computation
is critical to the convergence of the iterative algorithm. To obtain
a good initial value, the suboptimal estimator estimates $S_{m_i}$
by assuming it to be uniformly distributed. Furthermore, since the
estimate quality of the first stage is available, we use BLUE to
obtain $\hat{\theta}$ for exploiting the quality information,
instead of using the MLE in the M-step as in the standard EM
algorithm.
\par

\par

During the first stage of the iterative computation, the suboptimal
algorithm estimates $S_{m_i}$ under MMSE criterion.
This estimator requires \emph{a priori} probability of $S_{m_i}$
which depends on the unknown parameter $\theta$. The initial
distribution of $S_{m_i}$ is set to be uniform distribution. After
obtained a temporary estimate of $\theta$, we can apply it to update
the \emph{a priori} probability of $S_{m_i}$ and estimate $S_{m_i}$
iteratively. The MMSE estimator during the first stage is as
follows,
\begin{eqnarray}
\hat{S}_{m_i}&=& \mathrm{E}[S_{m_i}|h_i, \mat{y}_i]\nonumber\\
&=& \sum_{m_i=0}^{M}p(S_{m_i}|h_i, \mat{y}_i)S_{m_i}\nonumber\\
&=& \frac{\displaystyle\sum_{m_i=0}^{M}p(\mat{y}_i|h_i, S_{m_i})\hat{p}(S_{m_i})S_{m_i}}
{\displaystyle\sum_{m_i=0}^{M}p(\mat{y}_i|h_i,S_{m_i})\hat{p}(S_{m_i})}, \label{eq:mmse}
\end{eqnarray}
where $p(\mat{y}_i|h_i,S_{m_i})$ equals $p(\mat{y}_i|h_i,\mat{c}_m)$ in (\ref{eq:defpyihicm}),
and $\hat{p}(S_{m_i})$ is the estimate for \emph{a priori} PDF of $S_{m_i}$.
Obtained $\hat{\theta}$, we use  $p_A(S_{m_i}|\hat{\theta})$
to update $\hat{p}(S_{m_i})$, i.e., let
$\hat{p}(S_{m_i})=p_A(S_{m_i}|\hat{\theta})$. In this section, we
omit $\theta$ in $\hat{p}(S_{m_i})$ for notational simplicity,
though it depends on $\theta$.
\par

Now we derive the mean and variance of $\hat{S}_{m_i}$, which will be used in the BLUE of $\theta$.
\par

If $\hat{p}(S_{m_i})$ equals to its true value,  the MMSE estimator in (\ref{eq:mmse}) is unbiased because,
{\setlength{\arraycolsep}{1pt}
\begin{eqnarray}
    \mathrm{E}[\hat{S}_{m_i}|h_i]&=& \int_{\mathbb{C}^L}\hat{S}_{m_i}p(\mat{y}_i|h_i)
    \mathrm{d}\mat{y}_i\nonumber\\
    &=&\int_{\mathbb{C}^L}\frac{\sum_{m=0}^{M-1}p(\mat{y}_i|h_i,S_{m_i})\hat{p}(S_{m_i})S_{m_i}}
    {\sum_{m=0}^{M-1}p(\mat{y}_i|h_i,S_{m_i})\hat{p}(S_{m_i})}
    \nonumber \\ &&
    \sum_{m=0}^{M-1}p(\mat{y}_i|h_i,S_{m_i})\hat{p}(S_{m_i})\mathrm{d}\mat{y}_i\nonumber\\
    &=& \int_{\mathbb{C}^L}\sum_{m=0}^{M-1}p(\mat{y}_i|h_i,S_{m_i})\hat{p}(S_{m_i})S_{m_i}\mathrm{d}\mat{y}_i\nonumber\\
    &=& \sum_{m=0}^{M-1}\hat{p}(S_{m_i})S_{m_i}\nonumber\\
    &=& \mathrm{E}[S_{m_i}|h_i].
    \label{eq:mmseunbias}
\end{eqnarray}}
\par

However, $\hat{p}(S_{m_i})$ in our algorithm is inaccurate since we
use $\hat{\theta}$ instead of $\theta$. The MMSE estimate may be
biased, but it is hard to obtain this bias in practical systems. We
regard the MMSE estimator as an unbiased estimate in our suboptimal
algorithm.
\par

Given $h_i$ and $\mat{y}_i$, the variance of the MMSE estimate can be derived as,
{\setlength{\arraycolsep}{1pt}
\begin{eqnarray}
\mathrm{Var}[\hat{S}_{m_i}|h_i,\mat{y}_i]&=&
\mathrm{E}[S_{m_i}^2|h_i, \mat{y}_i]-\mathrm{E}^2[S_{m_i}|h_i,\mat{y}_i]\nonumber\\
&=& \frac{\sum_{m_i=0}^{M}S_{m_i}^2p(\mat{y}_i|h_i, S_{m_i})\hat{p}(S_{m_i})}
{\sum_{m_i=0}^{M}p(\mat{y}_i|h_i,S_{m_i})\hat{p}(S_{m_i})}-\hat{S}_{m_i}^2.
\nonumber\\
\label{eq:varmmse}
\end{eqnarray}}
\par

Then the BLUE for estimating $\theta$ is,
{\setlength{\arraycolsep}{1pt}
\begin{eqnarray}
\hat{\theta}&=&
\left(\sum_{j=1}^N\frac{1}{\sigma_s^2+\mathrm{Var}[\hat{S}_{m_j}|h_j,\mat{y}_j]}\right)^{-1}
\nonumber\\ &&
\sum_{i=1}^{N}\frac{\hat{S}_{m_i}}{\sigma_s^2+\mathrm{Var}[\hat{S}_{m_i}|h_i,\mat{y}_i]}.
\label{eq:mmseblue}
\end{eqnarray}}
\par

The iterative algorithm can be summarized as follows:
\begin{enumerate}
    \item[S1)] Let $\hat{p}(S_{m_i})=1/M$ as the initial value.
    \item[S2)] Compute $\hat{S}_{m_i}$, $i=1,\cdots,N$, and its variance with (\ref{eq:mmse}) and (\ref{eq:varmmse}).
    \item[S3)] Substitute $\hat{S}_{m_i}$ and its variance to (\ref{eq:mmseblue}) to get $\hat{\theta}$.
    \item[S4)] Update $\hat{p}(S_{m_i})$ using $p_A(S_{m_i}|\hat{\theta})$.
    \item[S5)] Repeat step S2) $\sim$ S4) until the algorithm converges or
    a predetermined number of iterations is reached.
\end{enumerate}
\par

Note that this suboptimal algorithm differs from the one proposed in
\cite{AysalBarner08Constrained}, which applies the maximal \emph{a
posteriori} (MAP) criterion to detect binary observations of the
sensors, then uses the results as the true value of the observations
in the MLE derived in noise-free channels. Our suboptimal algorithm
inherits the structure of the MLE developed in fading channels,
which gives ``soft'' estimates of the quantized observations at
first, and combines them with a linear optimal estimator afterward.
By conducting these two stages iteratively, the estimation accuracy
 improves rapidly. Although the suboptimal algorithm may converge to
some local optimal solutions due to the non-convexity of the
original optimization problem, it still performs fairly well as will
be shown in the simulation results. The convergence of the algorithm
will be studied by simulations in Section \ref{sec:orth:simulation}.


\section{Special Cases of the MLEs}\label{sec:specialcases}

To gain some insights on the decentralized MLE, in this section, we
first study two special cases of the MLEs when either the
observations of the sensors or the communications are perfect. After
that, we discuss the form of the MLE with known CSI using two
extreme case quantization, which are the AF transmission
(infinitesimal quantization resolution) and the 1-bit quantization
(rough most resolution). This will provide the connections of the
derived MLE with existing well-studied optimal estimators in these
special cases.
\par

\subsection{Ideal Observations and Ideal Communications}

Considering the approximate expression of the PMF shown in
(\ref{eq:psmtheta}), the likelihood equation with known CSI is 
approximated as,
{\setlength{\arraycolsep}{1pt}
\begin{eqnarray}
    \log p(\mat{Y}|\mat{h},\theta)
    &\approx& \sum_{i=1}^N\left(
    \log\left( \sum_{m=0}^{M-1}
    \exp\left(
    -\frac{\|\mat{y}_i-\sqrt{\mathcal{E}_d}h_i\mat{c}_m\|^2}{\sigma_c^2}-
    \right.\right.\right.\nonumber\\ && \left.\left.\left.
    \frac{(S_m-\theta)^2}{2\sigma_s^2}
    \right)
    \right)
    \right).
    \label{eq:quanlfo}
\end{eqnarray}}

This tells us that the MLE exploits both the signal level
information $\mat{y}_i$ and the data level information $S_m$ when
the quantization interval is small enough. If the observations are
perfect, say, no observation noises, we will show that the MLE
degenerates into a signal level optimal combiner---MRC followed by
data demodulation and parameter reconstruction. On the other hand,
if the communications are perfect, we will show that the MLE reduces
to a data level optimal fusion estimator---BLUE.

When CSI is unknown, we draw similar conclusions, except that the
MLE becomes a subspace-based estimator followed by data demodulation
and parameter reconstruction when there are no observation noises.

\subsubsection{Noiseless Observations}

When the observations of the sensors are ideal, i.e.,
$x_i=\theta$, $\forall i=1,\cdots,N$, we have,
\begin{equation}
  p(x|\theta)=\delta(x-\theta),
  \label{eq:pxthetaio}
\end{equation}
where $\delta(x)$ is the Dirac-delta function.
\par

We first consider the MLE with known CSI. Substituting
(\ref{eq:pxthetaio}) to (\ref{eq:lfsecond}) and ignoring all terms
that do not affect the estimation, the log-likelihood function is
simplified as,
\begin{equation}
  \log p(\mat{Y}|\mat{h},\theta) = -\sum_{i=1}^N\frac{\|\mat{y}_i-\sqrt{\mathcal{E}_d}h_i\mat{c}(\theta)\|_2^2}{\sigma_c^2}.
  \label{eq:lfio}
\end{equation}
\par

Since $\mat{c}(\theta)$ is a piecewise constant function that is not
derivable, we cannot compute the partial derivative of
(\ref{eq:lfio}) with respect to $\theta$. Instead, we first
regard $\mat{c}(\theta)$ as the parameter under estimated and
obtain the MLE for estimating $\mat{c}(\theta)$.
The log-likelihood function in (\ref{eq:lfio}) is concave with respect to
$\mat{c}(\theta)$ and its only maximum can be obtained
by solving the equation $\partial\log p(\mat{Y}|\mat{h},\theta)/\partial\mat{c}(\theta)=0$,
which is,
\begin{equation}
  \hat{\mat{c}}(\theta)=\frac{1}{\sqrt{\mathcal{E}_d}\sum_{j=1}^N|h_i|^2}\sum_{i=1}^N
  h_i^*\mat{y}_i.
  \label{eq:hatmatc}
\end{equation}
\par

Then we can use it as a decision variable to detect the transmitted
symbols and reconstruct $\theta$ according to the quantization rule
with the detection results.
\par

It shows that when the observations are perfect, the structure of the
MLE is the MRC concatenated with data demodulation and
parameter reconstruction. This is no surprise since in this case the
signals transmitted by different sensors are all identical, the
receiver at the FC is able to use the traditional diversity
technology to reduce the communication errors. Meanwhile, it is
unnecessary to use the redundant observations for data fusion.
\par

We then consider the MLE with unknown CSI. Upon
substituting (\ref{eq:pxthetaio}) to (\ref{eq:lfnocsifirst}) and
ignoring all terms that do not affect the estimation, the
log-likelihood function becomes,
\begin{eqnarray}
    \log p(\mat{Y}|\theta)&=&  \sum_{i=1}^N|\mat{y}_i\ctrans\mat{c}(\theta)|^2 \nonumber\\
    &=& \mat{c}(\theta)\ctrans\left( \sum_{i=1}^N\mat{y}_i\mat{y}_i\ctrans \right)\mat{c}(\theta).
    \label{eq:lfnocsiio}
\end{eqnarray}
\par

Again, we regard $\mat{c}(\theta)$ as the parameter to be estimated.
Recall that the energy of $\mat{c}(\theta)$ is normalized. Then the
problem that finds $\mat{c}(\theta)$ to maximize
(\ref{eq:lfnocsiio}) is a solvable quadratically constrained
quadratic program (QCQP)\cite{cvx},
\begin{eqnarray}
    \max_{\mat{c}(\theta)} && \mat{c}(\theta)\ctrans\left( \sum_{i=1}^N\mat{y}_i\mat{y}_i\ctrans \right)\mat{c}(\theta)
    \nonumber\\
    \mathrm{s.t.} && \|\mat{c}(\theta)\|_2^2=1.
    \label{eq:qcqp}
\end{eqnarray}
\par

The solution of (\ref{eq:qcqp}) can be obtained following the
results about QCQP in \cite{cvx},
\begin{equation}
    \hat{\mat{c}}(\theta)=\mat{v}_{max}\left(  \sum_{i=1}^N\mat{y}_i\mat{y}_i\ctrans  \right),
    \label{eq:vmax}
\end{equation}
where $\mat{v}_{max}(\mat{M})$ is the eigenvector corresponding to
the maximal eigenvalue of the matrix $\mat{M}$.
\par

This shows that when CSI is unknown at the FC in the case of
noise-free observations, the MLE becomes a subspace-based
estimator.

\subsubsection{Noiseless Communications}

When $\sigma_c^2\rightarrow 0$, we have
$\mat{y}_i=\sqrt{\mathcal{E}_d}h_i\mat{c}_{m_i}$. It means that
$\mat{y}_i$ is merely decided by $\mat{c}_{m_i}$, or equivalently
decided by $S_{m_i}$. Then the log-likelihood function becomes a
function of the quantized observation $S_{m_i}$.
\par

We first consider the MLE with known CSI. The
log-likelihood function with ideal communications is,
{\setlength{\arraycolsep}{1pt}
\begin{eqnarray}
  && \log p(\mat{Y}|\mat{h},\theta)\rightarrow \log p(\mat{S}|\mat{h},\theta)\nonumber\\
  &=& \sum_{i=1}^N\left(
  Q\left( \frac{S_{m_i}-\frac{\Delta}{2}-\theta}{\sigma_s} \right)-Q\left( \frac{S_{m_i}+\frac{\Delta}{2}-\theta}{\sigma_s} \right)
  \right),
  \label{eq:lfic}
\end{eqnarray}}
where $\mat{S}=[S_{m_1},\cdots,S_{m_N}]\trans$.
\par

Computing the derivative of (\ref{eq:lfic}), we have the likelihood
equation,
\begin{equation}
  \sum_{i=1}^N\frac{
  \exp\left(
  - \frac{(S_{m_i}-\frac{\Delta}{2}-\theta)^2}{2\sigma_s^2}
  \right)
  -
  \exp\left(
  - \frac{(S_{m_i}+\frac{\Delta}{2}-\theta)^2}{2\sigma_s^2}
  \right)
  }{ Q\left( \frac{S_{m_i}-\frac{\Delta}{2}-\theta}{\sigma_s} \right)-Q\left( \frac{S_{m_i}+\frac{\Delta}{2}-\theta}{\sigma_s} \right)}=0.
  \label{eq:lfeqic}
\end{equation}
\par

Generally, this likelihood equation has no closed-form solution.
Nonetheless, the closed-form solution can be obtained when the
quantization noise is very small, i.e., $\Delta\rightarrow
0$. Under this condition, $S_{m_i}\rightarrow x_i$ and
(\ref{eq:lfeqic}) becomes,
\begin{equation}
  \lim_{\Delta\rightarrow 0}\frac{\partial \log p(\mat{S}|\mat{h},\theta)}{\partial\theta}=\sum_{i=1}^N \frac{x_i-\theta}{\sigma_s^2}=0.
  \label{eq:lfeqicsec}
\end{equation}
\par

The MLE obtained from (\ref{eq:lfeqicsec}) is,
\begin{equation}
  \hat{\theta}=\frac{1}{N}\sum_{i=1}^Nx_i.
  \label{eq:lfblue}
\end{equation}
\par

It is also no surprise to see that the MLE reduces to BLUE,
which is often applied in centralized
estimation\cite{jinjun06power}, where the FC can obtain all raw
observations of the sensors.
\par

When the CSI is unknown at the FC, the receiver of the FC can
recover the quantized observations of the sensors with error-free if the
proper codebook, which will be discussed in Section \ref{sec:orth:codebookdesign},
is applied. Then the MLE with
unknown CSI also degenerates into the BLUE shown in
(\ref{eq:lfblue}). This is reasonable since only the structure of
communication depends on the channel information.
\par

The special cases of the MLEs with noiseless observations
or noiseless communications are summarized in Table
\ref{tab:spcialcase}.
\begin{table}[ht]
    \centering
    \caption{The special cases of the MLEs with known and unknown CSI.}
        \begin{tabular}{|c|p{1in}|p{1in}|}
        \hline MLE & Noiseless Obser. & Noiseless Comm. \& $\Delta\rightarrow 0$ \\ \hline\hline
        Known CSI & MRC & BLUE \\ \hline
        Unknown CSI & Subspace-based estimator & BLUE (proper codebook) \\ \hline
    \end{tabular}
    \label{tab:spcialcase}
\end{table}

\subsection{AF Transmission and Binary Quantization}

\subsubsection{AF Transmission}

Although the estimators we derived until
now consider digital communications, they can also be applied when
using the AF transmission, because the messaging function we
introduced in Section \ref{sec:orth:model} can describe the AF
transmission as well.
\par

The messaging function for AF transmission is,
\begin{equation}
    \mat{c}(x)=\alpha x,
    \label{eq:amcx}
\end{equation}
where $\alpha$ is the amplification gain. Since $\mat{c}(x)$ reduces
to a scalar, we rewritten it as $c(x)$.
\par

For AF transmission, we rewrite the energy normalization condition
as $\mathrm{E}[\mat{c}(x)\ctrans\mat{c}(x)]=1$. When
$\alpha=1/\sqrt{\mathrm{E}[\theta^2]+\sigma_s^2}$, this condition is
satisfied. Because $\theta$ is an unknown non-random parameter, we
cannot obtain $\mathrm{E}[\theta^2]$. To solve this problem, we
assume that $\theta$ is a random variable uniformly distributed in
$[-V,+V]$, then we have $\mathrm{E}[\theta^2]=V^2/3$. \footnote{By
this way, the message function $\mat{c}(x)$ can also be used to
present other modulations such as quadrature amplitude modulation
(QAM). }
\par

The received signal at the FC is,
\begin{eqnarray}
    y_i&=& \sqrt{E}_dh_ic(x_i)+n_{c,i}\nonumber\\
    &=& \sqrt{E}_dh_i\alpha (\theta+n_{s,i})+n_{c,i}.
    \label{eq:amyi}
\end{eqnarray}
\par

Substituting (\ref{eq:amcx}) and (\ref{eq:amyi}) to the
log-likelihood function with known CSI shown in (\ref{eq:lffinal}),
we can obtain the MLE with AF transmission. The derivation to obtain
the MLE in this way is rather involved due to the cross-correlation
between the real and imaginary parts of the received signals. In the
following, we will give an alternative derivation that is simple.
\par

We first find a vector of sufficient statistic, then derive the
log-likelihood function using this vector as an observation vector.
When $h_1,\cdots,h_N$ is known at the receiver,
it is not hard to show that $\mat{Y}_h=[h_1^*y_1,\cdots,h_N^*y_N]$
is a sufficient statistic of $\mat{Y}=[y_1,\cdots, y_N]$ for
estimating $\theta$, where,
\begin{equation}
    h_i^*y_i=\sqrt{\mathcal{E}_d}|h_i|^2\alpha(\theta+n_{s,i})+h_i^*n_{c,i}, \quad i=1,\cdots,N.
    \label{eq:defhiyi}
\end{equation}
\par

The real and imaginary parts of $h_i^*y_i$ are statistically
independent and Gaussian distributed. We can find that the mean and
variance of the real part is
$\sqrt{\mathcal{E}_d}|h_i|^2\alpha\theta$ and
$\mathcal{E}_d|h_i|^4\alpha^2\sigma_s^2 +|h_i|^2\sigma_c^2/2$, and
 the mean and
variance of the imaginary part is zero and $|h_i|^2\sigma_c^2/2$.
Ignoring the constant not associated with the MLE, we can obtain the
log-likelihood function as,
{\setlength{\arraycolsep}{1pt}
\begin{eqnarray}
    \log p\left( \mat{Y}_h |\theta \right)&=& 
    -\sum_{i=1}^N\left(
    \frac{(\Re\{h_i^*y_i\}-\sqrt{\mathcal{E}_d}|h_i|^2\theta)^2}
    {|h_i|^2\sigma_c^2+2\mathcal{E}_d\alpha^2|h_i|^4\sigma_s^2}+
    \right.\nonumber\\ && \left.
    \frac{(\Im\{h_i^*y_i\})^2}{|h_i|^2\sigma_c^2}
    \right),
    \label{eq:paflf}
\end{eqnarray}}
and the likelihood equation as,
{\setlength{\arraycolsep}{1pt}
\begin{eqnarray}
    \frac{\partial\log p(\mat{Y}_h|\theta)}{\partial\theta}&=& 
    \sum_{i=1}^N\left(\frac{2\sqrt{\mathcal{E}_d}\alpha\Re\{h_i^*y_i\}}
    {\sigma_c^2+2\mathcal{E}_d\alpha^2|h_i|^2\sigma_s^2}-
    \right.\nonumber\\ && \left.
    \frac{2\mathcal{E}_d\alpha^2|h_i|^2\theta}
    {\sigma_c^2+2\mathcal{E}_d\alpha^2|h_i|^2\sigma_s^2}
    \right)=0,
    \label{eq:partial}
\end{eqnarray}}
where $\Re\{z\}$ and $\Im\{z\}$ are the real and imaginary part of
$z$, respectively.
\par

The log-likelihood function has only one maximum that can be obtained
by solving the likelihood equation,
\begin{equation}
    \hat{\theta}=\frac{\displaystyle\sum_{i=1}^{N}\frac{\Re\{h_i^*y_i\}}
    {\sigma_c^2+2\mathcal{E}_d\alpha^2|h_i|^2\sigma_s^2}}
    {\displaystyle\sum_{i=1}^N\frac{\sqrt{\mathcal{E}_d}\alpha|h_i|^2}
    {\sigma_c^2+2\mathcal{E}_d\alpha^2|h_i|^2\sigma_s^2}},
    \label{eq:mleaf}
\end{equation}
which can degenerate to the optimal estimator proposed in
\cite{esdiversity07cui} under the assumptions therein.
\par

The asymptotic performance of the AF transmission in fading
orthogonal MACs are analyzed in
\cite{esdiversity07cui} and \cite{Bai2009outage}.

\subsubsection{Binary Quantization}

Considering the stringent constraint on the bandwidth of WSNs, many
contributions assume that the sensors use binary quantizer as the
local processor.
Our estimators can apply when binary quantizer is used. Based on the
system models in Section \ref{sec:orth:model}, the messaging
function with binary quantizer and binary phase-shift keying (BPSK)
modulation is
\begin{equation}
    \mat{c}_1(x)=\left\{
    \begin{array}{l l}
        -1, x \leq \tau\\
        +1, x > \tau
    \end{array}
    \right.,
    \label{eq:ctonebit}
\end{equation}
where $\tau$ is the quantization threshold, which equals to 0 for
the uniform quantizer we considered.
\par

Substituting $\mat{c}_1(x)$ to the likelihood function shown in
(\ref{eq:lffinal}), we have,
{\setlength{\arraycolsep}{1pt}
\begin{eqnarray}
    p(\mat{Y}|\mat{h}, \theta)
    &=&\prod_{i=1}^N\log\left(
    \exp\left( -\frac{|\mat{y}_i+\sqrt{\mathcal{E}_d}h_i|^2}{\sigma_c^2} \right)F_{n_s}(\tau-\theta)+
    \right.\nonumber\\ &&\left.
    \exp\left( -\frac{|\mat{y}_i-\sqrt{\mathcal{E}_d}h_i|^2}{\sigma_c^2} \right)(1-F_{n_s}(\tau-\theta))
    \right),\nonumber\\
    \label{eq:lfonebie}
\end{eqnarray}}
where $F_{n_s}(x)$ is the cumulative distribution function (CDF) of
the observation noise.
\par

Define
{\setlength{\arraycolsep}{1pt}
\begin{eqnarray}
    a(y_i)&=&
    \exp\left( -\frac{|\mat{y}_i+\sqrt{\mathcal{E}_d}h_i|^2}{\sigma_c^2} \right)-
    \exp\left( -\frac{|\mat{y}_i-\sqrt{\mathcal{E}_d}h_i|^2}{\sigma_c^2} \right),\nonumber\\
    \label{eq:defabya}
\end{eqnarray}}
and
{\setlength{\arraycolsep}{1pt}
\begin{eqnarray}
    b(y_i)&=&
    \exp\left( -\frac{|\mat{y}_i-\sqrt{\mathcal{E}_d}h_i|^2}{\sigma_c^2} \right)(1-F_{n_s}(\tau-\theta)),
    \label{eq:defaby}
\end{eqnarray}}
then (\ref{eq:lfonebie}) can be simplified as
\begin{equation}
    p(\mat{Y}|\mat{h},\theta)= \prod_{i=1}^N[a(y_i)F_{n_s}(\tau-\theta)+b(y_i)].
    \label{eq:lfonebitfinal}
\end{equation}
\par

This is the same as the likelihood function shown in (9) of
\cite{AysalBarner08Constrained} except for the presence of the channel
coefficients since we consider fading channels.
\par

\section{Discussions} \label{sec:DISCUSS}

\subsection{Transmission Codebook Issues}\label{sec:orth:codebookdesign}
When digital communications are used, the transmission codebooks can
represent various quantization, coding and modulation schemes. Here
we discuss the impact of the codebooks on the decentralized MLEs.
\par

We rewrite the conditional PDF with known CSI shown in
(\ref{eq:pyixi}) as,
{\setlength{\arraycolsep}{1pt}
\begin{eqnarray}
    p(\mat{y}_i|h_i,x)&=& \frac{1}{(\pi\sigma_c^2)^L}\exp\left(
    \commpart
    \right)\nonumber\\
    &=& \frac{1}{(\pi\sigma_c^2)^L}\exp\left( -\frac{\mathcal{E}_d|h_i|^2}{\sigma_c^2} \right)
    \nonumber\\ &&
    \exp\left(
    -\frac{\|\mat{y}_i\|_2^2}{\sigma_c^2}+\frac{2\sqrt{\mathcal{E}_d}\Re\{h_i\mat{y}_i\ctrans\mat{c}(x)\}}{\sigma_c^2}
    \right).
    \label{eq:pyixinew}
\end{eqnarray}
\par

Comparing the conditional PDF with unknown CSI $p(\mat{y}_i|x)$
shown in (\ref{eq:pdfyifinal}) with $p(\mat{y}_i|h_i, x)$ shown in
(\ref{eq:pyixinew}), we see that both PDFs depend on the correlation
between the received signals $\mat{y}_i$ and the transmitted symbols
$\mat{c}(x)$. With known CSI, the optimal estimator is a coherent
algorithm, since (\ref{eq:pyixinew}) relies on the real part of the
correlation, $\mat{y}_i\ctrans\mat{c}(x)$. With unknown CSI, the
optimal estimator is a non-coherent algorithm, since
(\ref{eq:pdfyifinal}) depends on the square norm of
$\mat{y}_i\ctrans\mat{c}(x)$. Because
$\mat{y}_i\ctrans\mat{c}(x)=\sqrt{\mathcal{E}_d}h_i^*\mat{c}\ctrans(x_i)\mat{c}(x)
+\mat{n}_{c,i}\ctrans\mat{c}(x)$, both MLEs depend on the
cross-correlation of the transmission symbols, $\mat{c}\ctrans(x_i)\mat{c}(x)$.
\par

Taking digital communications as an example,
if there exist two transmission symbols $\mat{c}_m$ and $\mat{c}_n$ in the
transmission codebook which have the same norm,
\begin{equation}
    \mat{c}_m=\mat{c}_ne^{j\phi},
\end{equation}
then $p(\mat{y}_i|x)$ will have two identical extrema since the MLE
with unknown CSI only depends on
$|\mat{y}_i\ctrans\mat{c}(x)|^2$. Such a phase ambiguity will lead
to severe performance degradation to the decentralized estimator.
Therefore, the auto-correlation matrix of the codebook plays a
critical role on the performance of the MLE, especially
when CSI is unknown.
\par

Many transmission schemes have this phase ambiguity problem. For
example, when the natural binary code and BPSK modulation are
applied to represent each quantized observation and to transmit, for
any $\mat{c}_m$ in such a transmission codebook, defined as
$\mat{C}_{tn}$, there exists $\mat{c}_{m'}$ in $\mat{C}_{tn}$ that
satisfies $\mat{c}_{m'}=-\mat{c}_m$. Therefore, $\mat{C}_{tn}$ is
not a proper codebook.

In order to cope with the phase ambiguity problem inherent in the
codebook $\mat{C}_{tn}$, we can simply insert training symbols into
the transmission symbols. Though heuristic, this approach can
provide fairly good performance because the MLE can exploit
the training symbols to estimate the channel coefficients implicitly
as we have shown.
\par

Since the MLEs are associated with the auto-correlation matrix of
the transmission codebook, this allows us to enhance the performance of the
estimators by systematically designing the codebook. Nonetheless,
this is out of the scape of this paper. Some preliminary results for
optimizing the transmission codebooks are shown in
\cite{myicassp09}.

\subsection{Asymptotic Performance of the MLEs with respect to $N$}\label{sec:orth:asyn}

We first consider the Cram\'er-Rao lower bound (CRLB) when CSI is unknown at the FC,
which is,
{\setlength{\arraycolsep}{1pt}
\begin{eqnarray}
    \mathrm{Var}[\hat{\theta}]&\geq& \left(
    -\mathrm{E}\left[
    \frac{\partial^2 \log p(\mat{Y}|\mat{h},\theta)}{\partial \theta^2}
    \right]
    \right)^{-1}\nonumber\\
    &=& \left(
    \sum_{i=1}^N\int_{\mathbb{C}^{L}}\frac{
    \left(\sum_{m=0}^{M-1}p(\mat{y}_i|\mat{c}_m)p'(S_m|\theta)\right)^2}
    {\sum_{m=0}^{M-1}p(\mat{y}_i|\mat{c}_m)p(S_m|\theta)}
    -
    \right.\nonumber\\ && \left.
    \sum_{m=0}^{M-1}p(\mat{y}_i|\mat{c}_m)p''(S_m|\theta)\mathrm{d}\mat{y}_i
    \right)^{-1}\nonumber\\
    &=& \frac{1}{N}\left(\int_{\mathbb{C}^L}
    \frac{\left(\sum_{m=0}^{M-1}p(\mat{y}|\mat{c}_m)p'(S_m|\theta)\right)^2}
    {\sum_{m=0}^{M-1}p(\mat{y}|\mat{c}_m)p(S_m|\theta)}\mathrm{d}\mat{y}-
    \right.\nonumber\\ && \left.
    \sum_{m=0}^{M-1}p''(S_m|\theta)
    \right)^{-1},
    \label{eq:crlb}
\end{eqnarray}}
where $p'(S_m|\theta)$ and $p''(S_m|\theta)$ is the first and second order partial derivatives
with respect to $\theta$, respectively.
\par

It shows that the CRLB of the MLE with unknown CSI
decreases with the factor of $1/N$, which is the same as the BLUE
lower bound of the centralized estimation\cite{jinjun06power}. This
is due to the fact that given $\theta$, the received signals
$\mat{y}_i$ from different sensors are statistically identical
distributed and independent among each other. Therefore, all these
signals contribute equally for reducing the estimation errors.
\par

When CSI is available at the FC, given $h_i$, the received signals
are no longer identical distributed. In this case, the CRLB will
depend on the channel realization, which is very hard to derive.
However, since more information can be exploited for estimation, we
can infer that the CRLB with known CSI is always lower than that
with unknown CSI. In other words, the asymptotic performance of the
MLE with known CSI will be no worse than that of the MLE with
unknown CSI.
\par

\subsection{Computational Complexity}\label{sec:orth:compcomplex}

\subsubsection{MLE}

We take the MLE with known CSI as an example to analyze the
computational complexity. The analysis for the MLE with unknown CSI
is similar.
\par

The MLE can be found by performing exhaustive searching. In order to
make the MSE introduced by the discrete searching neglectable, we
let the searching step-size be less than $\Delta/N$,
%
Then we need to compute the value of the likelihood function at
least $M\times N$ times to obtain the MLE.
\par

The FC applies (\ref{eq:lfd}), (\ref{eq:defpyihicm}), and
(\ref{eq:defpsm}) to compute the values of the likelihood function
with different $\theta$. The exponential term in
(\ref{eq:defpyihicm}) is independent of $\theta$, thus it can be
computed before searching and be stored for future use.
\par

Given $\theta$, we still need to compute $p(S_m|\theta)$,
$m=0,\cdots,M-1$, which complexity is $O(M)$, then to conduct $M$
additions and $M$ multiplications to obtain each value of the
likelihood function. Thus the computational complexity for getting
one value of $\log p(\mat{Y}|\mat{h},\theta)$ is $O(MN)$.
\par

After considering the operations required by the exhaustive
searching, the overall complexity of the MLE is $O(M^2N^2)$.
\par

\subsubsection{Suboptimal Estimator}

The estimator presented in Section \ref{sec:orth:subopt} uses an
iterative algorithm. For each iteration, we need to get
$\hat{S}_{m_i}$ and its variance with (\ref{eq:mmse}) and
(\ref{eq:varmmse}), then obtain the estimate of $\theta$ with
(\ref{eq:mmseblue}). The complexity is similar to that of computing
the log-likelihood function, which is $O(MN)$.
\par

If the algorithm converges after $I_t$ times iteration, the
complexity of the suboptimal estimator will be $O(I_tMN)$.
\par

%

\section{Simulations}\label{sec:orth:simulation}

We use the MSE of estimating $\theta$ as the metric to evaluate the
performance of the estimators. The observation SNR considered in
simulations is defined as\cite{jjxiao05decentralized},
\begin{equation}
    \gamma_s=20\log_{10}\left( \frac{W}{\sigma_s} \right).
\label{eq:nmobsersnr}
\end{equation}
\par

We use $\mathcal{E}_d$, the energy consumed by each sensor to
transmit one observation, to define the communication SNR in order
to fairly compare the energy efficiency of the estimators with
different transmission schemes. The communication SNR is then,
\begin{equation}
    \gamma_c=10\log_{10}\left( \frac{\mathcal{E}_d}{N_0} \right).
\label{eq:nmcommsnr}
\end{equation}
\par

The codebooks used in the simulations are summarized in Table
\ref{tab:codebook}. Consider the general features of WSNs that
short data packets are usually transmitted and each sensor is of low
cost. We use a simple error-control coding scheme, cyclic redundancy
check (CRC) codes with generator polynomial $G(x)=x^4+x+1$,
as an example of the coded transmission. Its
codebook is denoted as $\mat{C}_{tc}$.
For comparison, uncoded transmission
is also evaluated, which codebook is denoted as $\mat{C}_{tn}$. We
consider BPSK modulation to generate all codebooks. Because the code
length of the uncoded transmission is shorter than that of the coded
transmission, the energy to transmit each symbol will be higher for
the given $\mathcal{E}_d$. Due to the phase ambiguity problem
discussed in Section \ref{sec:orth:codebookdesign}, we use the
codebook with training symbols (TS) $\mat{C}_{tp}$ whenever we evaluate
the estimators with unknown CSI unless specified.
\par

\begin{table}[htb]
    \centering
    \caption{The Summary of the Codebooks Considered.}
        \begin{tabular}{|c|c|c|c|}
        \hline Codebook & ECC & TS & Modulation \\ \hline\hline
        $\mat{C}_{tn}$ & No & No & BPSK \\ \hline
        $\mat{C}_{tc}$ & CRC & No & BPSK \\ \hline
        $\mat{C}_{tp}$ & No & Yes & BPSK\\ \hline
    \end{tabular}
    \label{tab:codebook}
\end{table}

Two estimators with ideal communications are shown as the baseline,
which MSEs can be served as the performance lower bound. They are
BLUE which MSE is $\sigma_s^2/N$, and Quasi-BLUE when considering
quantization noise\cite{jinjun06power}. The Quasi-BLUE bound is a
more practical lower bound for comparison since we consider
quantization for all estimators in this paper only except for the
estimator with AF transmission.

\subsection{Influence of the Quantization Bit-Rate}

We first examine the impact of quantization bit-rate of the sensors.
Three WSNs are considered, where the sensors in different WSNs use
different quantization bit-rates.
The three quantization bit-rates are set to be $K = 1$, 2, and 4, respectively.
The sensors apply $\mat{C}_{tn}$ as the transmission codebook, where the length
of the transmitted symbols
is $L= K$.
In this simulation, we let both the total energy and the total bandwidth
consumed by the networks to be identical when using different quantization levels.
Due to the total network bandwidth constraint,
the numbers of the active sensors when $K = 1$, 2, and 4
are 40, 20, and 10, respectively.
Due to the total network energy constraint, the energy consumed by each sensor
to transmit one observation is also different.
For example, if the transmitted energy of a sensor when $N=40$ is $\mathcal{E}_d$,
the transmitted energy is $2\mathcal{E}_d$ and $4\mathcal{E}_d$ when $N=20$ and 10,
respectively.
\par

\begin{figure}[thp]
    \begin{center}
        \includegraphics[width=\figwidth]{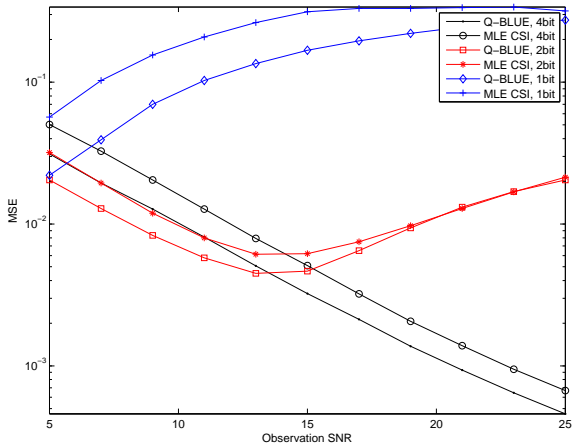}
    \end{center}
    \caption{The MSEs of the MLE with different $K$.
    The MSE of MLE with known CSI is marked as \emph{MLE CSI} in the legend, and Quasi-BLUE lower bound is marked as
    \emph{Q-BLUE}. The communication SNR is 6 $dB$ for 4 bit quantization, 3 $dB$ for 2bit quantization,
    and 0 $dB$ for binary quantization.}
    \label{fig:bit1}
\end{figure}

We compare the MSEs of the MLE with known CSI and the Quasi-BLUE
lower bound for different quantization bit-rates in Fig.
\ref{fig:bit1}. It is shown that low quantization bit-rate is only
applicable for the cases with extremely low observation SNR. At
medium and high observation SNR levels, the optimal estimator with 1
or 2-bit quantization is inferior to that with 4-bit quantization
under the same constraints of total energy and bandwidth.
\par

This indicates that we should design the quantization bit rate
according to the observation SNR to reduce the resource consumption
of the network. For high observation SNR, we should employ high
bit-rate with reduced number of the active sensors. We
note that similar conclusion is drawn in \cite{Li2009distributed}
except that \cite{Li2009distributed} considers error-free
communications whereas we consider the communications over fading
channels.
\par

\subsection{Convergence of the Suboptimal Estimators}
We then study the convergence of the suboptimal estimators.
Figure \ref{fig:iter} depicts the MSEs of the suboptimal estimators as a
function of the number of iterations. It is shown that the MSEs of
the suboptimal estimators will converge after two iterations at
different communication SNRs no matter if CSI is known.
\begin{figure}[thp]
\begin{center}
\includegraphics[width=\figwidth]{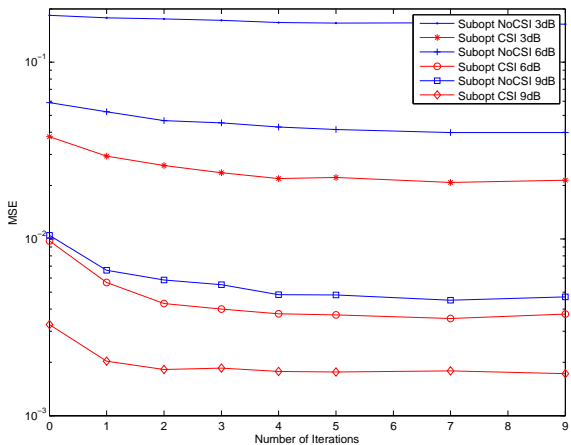}
\caption{The convergence of the suboptimal estimators when
$\gamma_s=20$ $dB$, $N=10$. In the legend, \emph{NoCSI} indicates
suboptimal estimator with unknown CSI, and \emph{CSI} stands for the
suboptimal estimator with known CSI. The communication SNRs are 3
$dB$, 6 $dB$ and 9 $dB$, which are marked in the legend. }
\label{fig:iter}
\end{center}
\end{figure}

\subsection{MSE versus the Communication SNR}
Figure \ref{fig:csi20db} depicts the MSEs of the estimators with
known CSI. Except for the estimator using AF transmission, all other
estimators use digital communications with a 4-bit uniform quantizer
($M=16$).
\par

To demonstrate the performance gain of the proposed estimators that
jointly optimize demodulation and parameter estimation, two
traditional fusion based estimators and a MRC based estimators are
simulated. In the fusion based estimators, the FC first demodulates
the transmitted data from each sensor, then reconstructs the
observation of each sensor from the demodulated symbols following
the rule of quantization, afterward combines these estimated
observations with BLUE fusion rule to produce the final estimate of
$\theta$. When ECCs are applied at the sensors, the receiver at the
FC will exploit its error detection ability to discard the data that
cannot pass the error check.
In the MRC based estimator, the FC
first combines the received signals from all sensors, then
demodulates the transmitted symbols. Finally the FC obtains the
estimate of $\theta$ using the detected symbols according to the
quantization rule.
\par

Except for the fusion based estimator with ECC uses codebook $\mat{C}_{tc}$,
all estimators use codebook $\mat{C}_{tn}$ in this simulation.
\par

\begin{figure}[thp]
\begin{center}
\includegraphics[width=\figwidth]{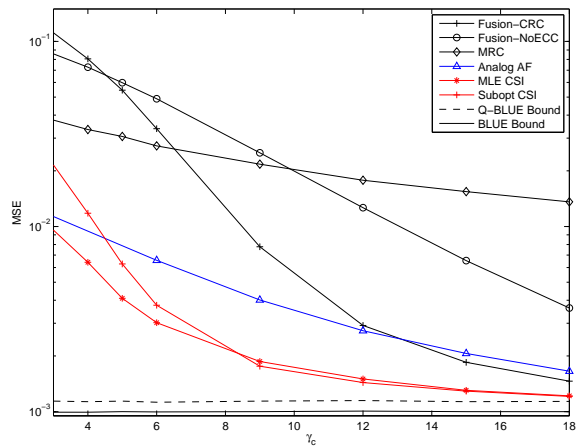}
\caption{The MSEs of the estimators with known CSI as a function of
communication SNR when $N=10$, $\gamma_s=20$ $dB$. In the legend,
``\emph{Fusion-CRC}'' and ``\emph{Fusion-NoECC}'' stand for two
fusion based estimators using the codebook $\mat{C}_{tc}$ and
$\mat{C}_{tn}$, ``\emph{MRC}'' stand for the MRC based estimators,
``\emph{Analog AF}'' stands for the MLE with AF transmission,
``\emph{MLE}'' and ``\emph{Subopt}'' denotes the MLE and suboptimal
estimators, respectively, and ``\emph{Q-BLUE Bound}'' and
``\emph{BLUE Bound}'' stands for two lower bounds with ideal
communications.} \label{fig:csi20db}
\end{center}
\end{figure}

It is shown that the MLE and suboptimal estimators outperform both
the MRC based and the fusion based estimators. The MSEs of the MLE
and suboptimal estimator approach the Quasi-BLUE lower bound
rapidly along with the increasing of the communication SNR, whereas
the suboptimal estimator degrades a little at low SNR. The MSE of
MLE using AF transmission is larger than that using digital
transmission, since AF transmission is no longer optimal in fading
channels.
\par

According to the performance analysis for BPSK modulation in
Rayleigh fading channels\cite{digitalcomm}, the BER of the
transmission scheme with codebook $\mat{C}_{tn}$ exceeds $0.15$ when
the communication SNR is lower than $3$ $dB$. ECC can improve the
transmission performance for high communication SNR, but it causes
more errors for low SNR. For the transmission schemes using CRC, the
BER is even worse because long codes will reduce the transmission
energy per symbol. For such a high BER, the fusion based estimators,
especially those with ECCs, do not perform well. Most of the
demodulated data will be dropped due to the error check,
thus the fusion estimators do not have enough information to
exploit, which finally leads to worse MSE performance.
\par

The performance of the estimator based on the MRC is much worse than
the proposed estimators, which shows the significant impact of the
observation noise.

\begin{figure}[thp]
\begin{center}
\includegraphics[width=\figwidth]{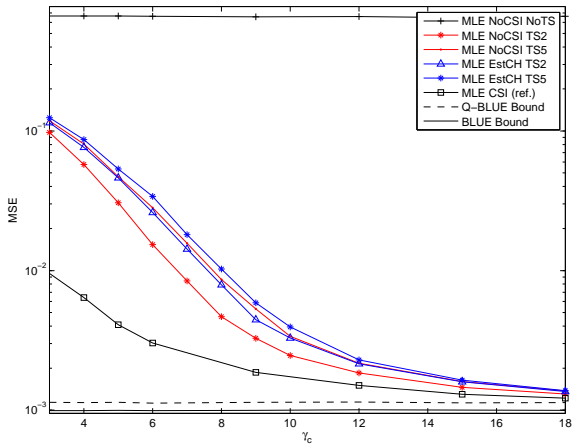}
\caption{The MSEs of the MLEs with different level of channel
information, where $N=10$ and $\gamma_s=20$ $dB$. In the legend,
``\emph{MLE NoCSI NoTS}'', ``\emph{MLE NoCSI TS2}'' and ``\emph{MLE
NoCSI TS5}'' stands for the MLE with unknown CSI and different
number of training symbols, respectively. ``\emph{MLE EstCH}''
denotes the MLE with known CSI that applies $\hat{h}_i$ as true
value of $h_i$. ``\emph{MLE CSI (ref.)}'' is the MLE with known CSI,
just for a reference.} \label{fig:nocsi20db}
\end{center}
\end{figure}

\begin{figure}[thp]
\begin{center}
\includegraphics[width=\figwidth]{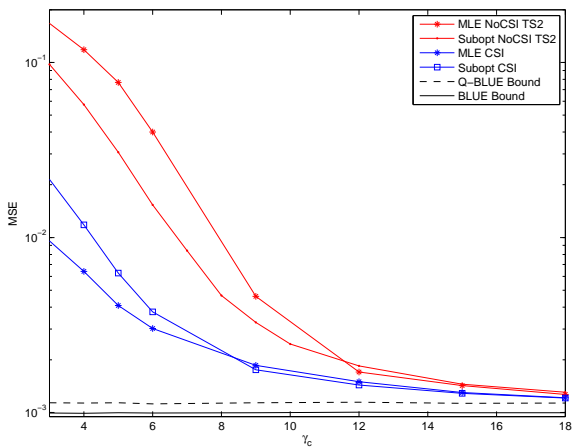}
\caption{The MSEs of the MLE and suboptimal estimators with training
symbols and with known CSI, where $N=10$ and $\gamma_s=20$ $dB$. In
the legend, \emph{MLE} and {Subopt} means MLE and suboptimal
estimators, respectively. } \label{fig:nocsisub}
\end{center}
\end{figure}

In Fig. \ref{fig:nocsi20db}, the MSEs of the MLEs with 
unknown CSI are shown. 
Two MLEs, which differently use training symbols, are considered.
One is the MLE with training
symbols as shown in (\ref{eq:lfsemifinal}), and the other is the
estimator as in (\ref{eq:lfeshfinal}), which uses the estimated
channel coefficients as their true values. Besides the codebook
$\mat{C}_{tn}$ without training symbols, we also evaluate the
codebooks with 2 and 5 training symbols. It is shown that if
$\mat{C}_{tn}$ is applied as the codebook for the MLE with unknown
CSI, the MLE exhibits rather high MSE that cannot be improved by
increasing the communication SNR. This validates our analysis in
Section \ref{sec:orth:codebookdesign} that the phase ambiguity of
$\mat{C}_{tn}$ will lead to severe performance degradation of the
estimator. When we insert training symbols, the performance of the
MLE with unknown CSI improves significantly, but it is still much
worse than that of the MLE with known CSI at low communication SNR
levels. It is interesting to see that using more training symbols do
not improve the performance of the MLE as expected. This is because
the energy for transmitting an observation is fixed, inserting
training symbols will reduce the energy for the data symbols. Our
simulations show that the best performance is obtained when $L_p=2$.
This is consistent with the observation of \cite{wmxicsp}, where the
optimal $L_p$ equals $\sqrt{K}$.
\par

To further observe the impact of different levels of CSI on the
optimal and suboptimal estimators, Fig. \ref{fig:nocsisub} shows the
MSEs of the MLE and suboptimal estimators with known CSI and with
unknown CSI but using two training symbols. Similar to the
estimators with known CSI, the suboptimal estimator with training
symbols is inferior to the MLE at low communication SNR. However,
the performance of the suboptimal estimator degrades less than the
MLE due to the channel estimation errors.
\par

\subsection{MSE versus the Number of Sensors}

\begin{figure}[thp]
\begin{center}
\includegraphics[width=\figwidth]{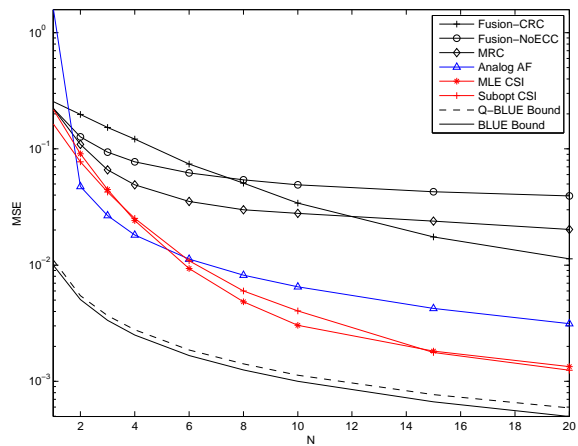}
\caption{The MSEs of the estimators with known CSI, where
$\gamma_c=6$ $dB$ and $\gamma_s=20$ $dB$. The meaning of the legends
is the same as Fig.\ref{fig:csi20db}. } \label{fig:csin}
\end{center}
\end{figure}

\begin{figure}[thp]
\begin{center}
\includegraphics[width=\figwidth]{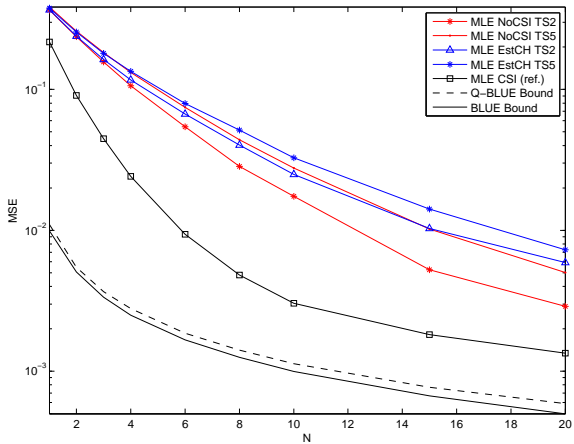}
\caption{The MSEs of the estimators with training symbols and with
estimated CSI when $\gamma_c=6$ $dB$ and $\gamma_s=20$ $dB$. The
legends are the same as Fig.\ref{fig:nocsi20db}. }
\label{fig:nocsin}
\end{center}
\end{figure}

Figure \ref{fig:csin} and Fig. \ref{fig:nocsin} show the MSEs of the
estimators with known CSI and unknown CSI as a function of
the number of sensors, $N$. We can see that the MSEs of all the
estimators decrease at the speed of $1/N$ for large enough $N$, but
the MSEs cannot approach the lower bound due to communication
errors. Compare the MSEs of the MLEs, we can see that the results
validates our asymptotic performance analysis for MLEs both with
known CSI and unknown CSI in \ref{sec:orth:asyn}. From Fig.
\ref{fig:csin}, we can observe that the proposed estimators perform
much better than the fusion based estimators and the MRC based
estimators. It means that the networks with traditional approaches
must activate more sensors to achieve the same MSE performance as
those with our estimators, which will lead to low energy and
bandwidth efficiency.

\subsection{Computational Complexity of the Estimators}

To evaluate the computational complexity, we record the time
consumed by 10,000 Monte-Carlo simulations for the proposed
estimators with known CSI. Table \ref{tab:fast} shows the
computation time in seconds at different communication SNR levels.
The step-size for the exhausting searching of the MLE is set to
$\Delta/N$. The number of the iterations for suboptimal estimator is
set to 2 according to the convergence analysis.
\par

It is shown that the computation time of the suboptimal estimator is
much less than that of the MLE, and is almost invariant with the
communication SNR since the number of the iterations is fixed. The
computation time consumed by the MLE varies slightly, which comes
from the implementation of the truncate exponential function in
simulation codes.

\begin{table}[htbp]
    \centering
    \caption{The computation time in seconds consumed by simulating the estimators with known CSI. }
    \label{tab:fast}
    \begin{tabular}{|l | r r r r r|}
        \hline
        \multicolumn{1}{|c|}{$\gamma_c$} &
        \multicolumn{1}{c}{3 dB} & \multicolumn{1}{c}{6 dB} & \multicolumn{1}{c}{9 dB} &
        \multicolumn{1}{c}{12 dB} & \multicolumn{1}{c|}{15 dB} \\ \hline\hline
        MLE & 274.12 & 270.86 & 258.16 & 234.56 & 207.58 \\
        Suboptimal & 6.80 & 6.88 & 6.78 & 6.58 & 6.35 \\
        \hline
    \end{tabular}
\end{table}

\par

\section{Conclusion}\label{sec:orth:con}

In this paper, we studied the decentralized estimation for a
deterministic parameter using digital communications over orthogonal
multiple-access fading channels with a uniform multiple-bit
quantizer. By introducing a general messaging function, the proposed
estimators can be applied for digital communication systems using
various quantization, coding and modulation schemes and for analog
communication systems such as those using the well-studied AF
transmission.
\par

We derived the MLEs with known and unknown CSI. When
inserting training symbols before the data symbols, the MLE with
unknown CSI estimates channels implicitly and exploits the channel
estimates in an optimal way. Following the structure of the MLE, we
designed a suboptimal estimator that has affordable complexity and
converges rapidly. It performs as well as the MLE at high
communication SNR and has minor performance loss at low
communication SNR.
\par

Simulation results show that both the MLEs and the suboptimal
estimators outperform the traditional MRC based and fusion based
estimators, and the estimators using digital communications
outperform those using AF transmission in Rayleigh fading channels.
Compared with the WSN that using binary quantization for
decentralized estimation, the system using multiple-bit quantization
has superior energy and bandwidth efficiency. Therefore, even with
the strict bandwidth constraints, we suggest that the WSNs should
use multiple-bit quantization rather than binary quantization when
the observation SNR is relative high.
\par

\bibliographystyle{IEEEtran}
\bibliography{ref/IEEEfull,ref/book,ref/des,ref/net}

\end{document}